\documentclass[a4paper]{article}

\usepackage{spconf,amsmath,graphicx,url,hyperref,color,multirow}

\title{Conversational End-to-End TTS for Voice Agents}
\name{Haohan Guo${^1}{^*}{^+}$\thanks{$*$ Work performed as an intern at Microsoft.}, Shaofei Zhang${^2}{^+}$\thanks{$+$ Equal contribution}, Frank K. Soong$^2$, Lei He$^2$, Lei Xie$^{1\dagger}$\thanks{$\dagger$ Corresponding author}}
\address{
$^1$Audio, Speech and Language Processing Group (ASLP@NPU), \\ School of Computer Science, Northwestern Polytechnical University, Xi’an, China \\
$^2$Microsoft China\\
\href{mailto:hhguo@nwpu-aslp.org}{\nolinkurl{{hhguo,lxie}@nwpu-aslp.org}, \nolinkurl{{frankkps, helei, shazh}@microsoft.com}}
}

\begin{document}

\maketitle
\begin{abstract}
End-to-end neural TTS has achieved excellent performance on reading style speech synthesis. However, it is still a challenge to build a high-quality conversational TTS due to the limitations of corpus and modeling capability. This study aims at building a conversational TTS for a voice agent under sequence to sequence modeling framework. We firstly construct a spontaneous conversational speech corpus well designed for the voice agent with a new recording scheme ensuring both recording quality and conversational speaking style. Secondly, we propose a conversation context-aware end-to-end TTS approach that employs an auxiliary encoder and a conversational context encoder to specifically reinforce the information about the current utterance and its context in a conversation as well. Experimental results show that the proposed approach produces more natural prosody in accordance with the conversational context, with significant preference gains at both utterance-level and conversation-level. Moreover, we find that the model has the ability to express some spontaneous behaviors like fillers and repeated words, which makes the conversational speaking style more realistic.
\end{abstract}


\section{Introduction}

Text to speech (TTS) has been playing an increasingly important role in human-machine conversation \cite{rojc2014tts}, enabling the machine to talk with users. However, the existing TTS technologies still cannot achieve satisfying performance and immersive experience in conversation-oriented tasks. It still desires more human-like natural speech with conversational speaking style adapting to specific conversational context. In order to build a high quality conversational TTS system, at least two problems need to be solved: 1) an effective method of developing a conversational speech corpus, and 2) a high performance TTS model of capturing rich prosody in conversations.

A standard corpus is usually composed of well-designed transcripts and high-quality recordings. This kind of corpus mostly requires the speaker to speak every single utterance in a consistent reading style. Hence it is difficult for the speaker to read context-aware conversational transcripts with natural conversation-style prosody under this recording scheme. In \cite{husin2011creating}, the corpus is built by letting two speakers freely discuss a topic to present real spontaneous conversational speaking style. But it also brings other problems, such as the unclear pronunciation, over-disfluent prosody and background noise, which may aggravate the difficulty of data annotation and modeling. To alleviate these problems, in this paper, we propose a new scheme for building spontaneous conversational corpus, which consists of three steps: scene and dialogue design, recording in the form of performance and transcribing. By combining the reading style recording scheme and the free talk scheme in \cite{husin2011creating}, it becomes more effective to build a conversation corpus with clear pronunciation, high audio quality and spontaneous conversational speaking style.

\begin{figure*}[!htp]
    \centering
    \includegraphics[width=0.9\textwidth]{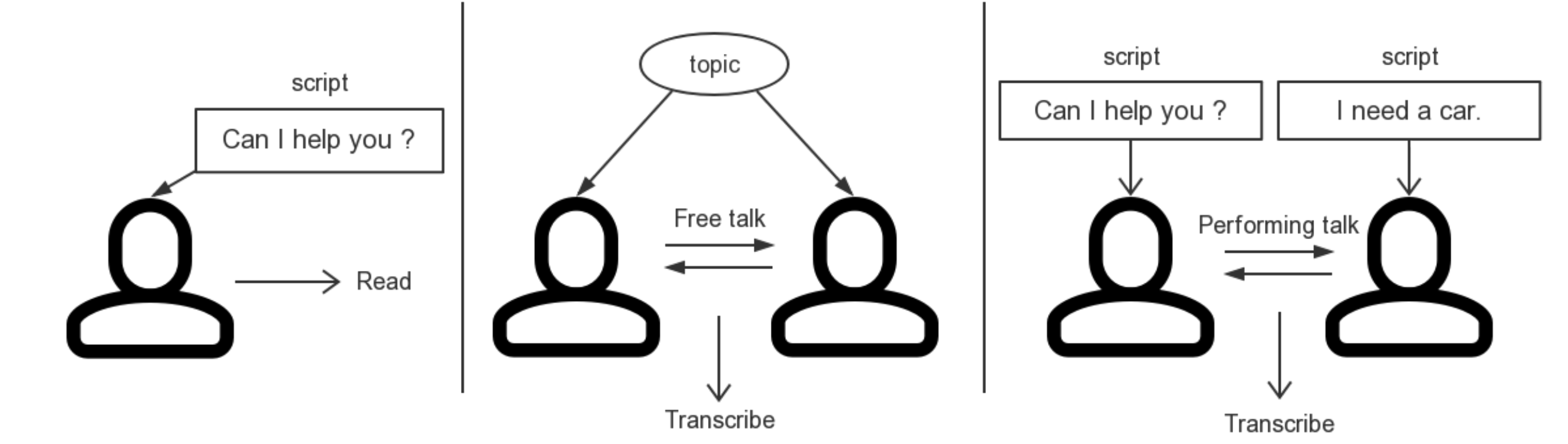} 
    \caption{The different recording approaches for TTS corpus}
    \label{fig:corpus}
\end{figure*}

Conversational corpus has the salient characteristics of high variety of prosody and strong context dependency. Moreover, due to the collection and annotation difficulty mentioned above, the corpus usually has a small size. We need rich text features and a high-performance model to build a conversational TTS system. Most previous researches are based on HMM-based or DNN-based statistical parametric speech synthesis (SPSS) \cite{koriyama2010conversational}, which utilize complicated conversation-related labels, e.g. speech or dialog acts \cite{syrdal2008dialog} and extended context \cite{koriyama2011use}, to directly provide rich textual information to compensate for its limited modeling capability. However, expensive labeling cost and insufficient representation of conversation make it more difficult to build a highly expressive conversational TTS. To avoid these problems, in this paper, we propose a new conversation context-aware TTS approach based on the state-of-the-art sequence-to-sequence (seq2seq) model. End-to-end TTS \cite{wang2017tacotron,shen2018natural,li2018close,DBLP:conf/interspeech/CongY0YW20,DBLP:conf/interspeech/YangYWWX20} based on seq2seq paradigm has recently shown great modeling capability that can synthesize natural speech directly from a character or phoneme sequence. This brings possibility to abandon conventional complicate labels. Besides, we introduce an auxiliary encoder to help to produce better prosody by extracting more useful latent semantic and syntactic features from BERT embedding and statistical features on the syntactic structure. Moreover, different from the conventional conversational TTS, we directly use a conversation context encoder to extract prosody-related information from the chat history, which is represented by a sequence of utterance-level BERT embedding.

This paper will firstly introduce our new recording scheme for spontaneous conversational speech corpus, and a Chinese voice agent corpus is developed in this way. Then we describe our proposed conversation context-aware end-to-end TTS system in detail, including the end-to-end model, auxiliary encoder and conversation context encoder. Finally, we conduct a CMOS test with a test set containing typical conversations between customer and agent, to evaluate the performance of our approach in both utterance level and conversation level. Experimental results show that both the auxiliary encoder and the conversational context encoder can effectively improve naturalness.
In addition, we find that the model has the ability to express some spontaneous behaviors, like fillers and repeated words, which leads to more realistic conversation speaking style.

\section{Spontaneous Conversational Corpus}

Conversational TTS requires not only high-quality recording and well-designed scripts, but also natural and realistic conversational scenes. In the paper, we adopt a new recording approach to meet these requirements. 

\subsection{Performing Recording}

For the standard TTS corpus, the speaker is often required to read an utterance fluently and accurately according to the well-designed transcripts, as shown in the most left part of Figure \ref{fig:corpus}. This method can ensure the high-quality recording \cite{syrdal2010speech}, but apparently lacks the variation of the speaking style matching the pre-designed text. It may not be suitable for dialogue scenes in which the context-aware prosody is highly desired. Another approach for conversational recording (the middle one in Figure \ref{fig:corpus}) is to make the speakers have task-oriented conversations \cite{maekawa2003corpus,mori2011constructing}. The transcripts of the free-talking are not pre-designed, which may lead to lower pronunciation clarity and over-disfluent prosody. All of these factors will aggravate the difficulty of data annotation and modelling. 

To keep both natural conversational speaking style and high recording quality, we propose a performing recording approach (the most right one in Figure \ref{fig:corpus}), which is composed of three steps.
\begin{itemize}
    \item Design conversational scenarios and transcripts. Most conversations are designed for voice agents, like scheduling, booking flight tickets and hotel rooms. The script design can better ensure the expected variety of content, sentence pattern, and conversational context.
    \item Speakers perform according to the scripts. They are allowed to modify part of the content and insert appropriate spontaneous behaviors for more natural conversational speaking style, such as fillers, repeated word, etc.
    \item Transcribing. We will transcribe the speaker's actual speech content to ensure correct pronunciation in the transcripts for modelling.
\end{itemize}
By allowing the speakers to perform freely within the well-designed conversational scripts and then transcribing the speech according to the actual speech content, we can effectively obtain a conversational TTS corpus with good recording quality and sufficient agent-orient conversational speaking style. In this way, we finally collect 45 conversations between two female native Chinese speakers, an agent and a customer (totally 6 hours, 3 hours per speaker). 

\subsection{The Spontaneous Behaviors}
\label{sec:Spontaneous}

There have been many studies for the spontaneous behavior, such as \cite{szekely2019casting,qader2018disfluency}. Usually, we focus on the most salient feature of it, \textit{disfluency}. In our corpus, we include the following behaviors:
\begin{itemize}
    \item Fillers, such as "um", "oh", "aha", "uh", etc.
    \item Repeated word or phrase. The word or phrase is spoken twice because of anxious or stammer.
    \item False start. The speaker sometimes interrupts the current sentence, then begins a new sentence.
    \item Reduced speech rate or pause due to the non-verbal behavior.
\end{itemize}

Different from using complex labels for the spontaneous behavior in SPSS, in this paper, we propose to use a conversation context-aware end-to-end TTS model to learn to express these behaviors by itself.

\begin{figure}[htp]
    \centering
    \includegraphics[width=0.45\textwidth]{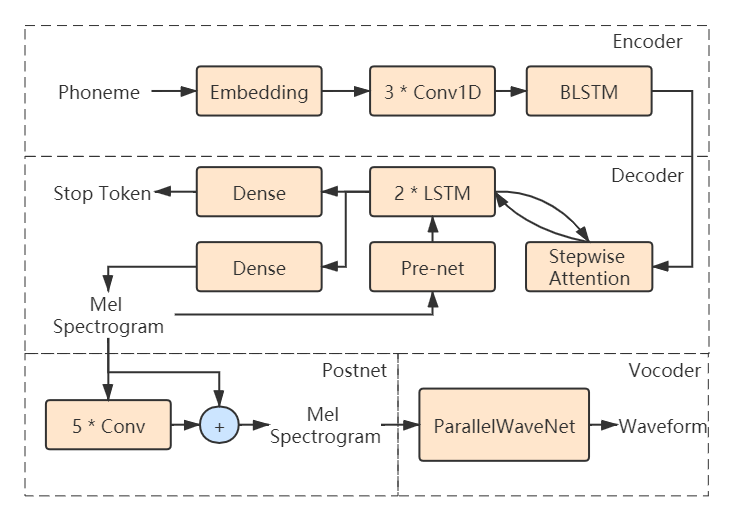}
    \caption{The model structure of the modified Tacotron2.}
    \label{fig:taco2}
\end{figure}

\section{Conversational Context-Aware End-to-End TTS}

The prosody in a conversation highly depends on both of the current text and its context. So a powerful model is desired to learn to extract rich prosody-related information from them. In this section, we introduce a conversational context-aware end-to-end TTS approach which can produce better conversational speech with the help of an auxiliary encoder and a conversational context encoder.

\subsection{End-to-End TTS}
\label{sec:baseline}

As shown in Figure \ref{fig:taco2}, our end-to-end TTS system is designed based on Tacotron2 \cite{shen2018natural}, which has shown excellent performance on naturalness and sound quality. It is an attention-based seq2seq mapping model composed of an encoder, an auto-regressive decoder, PostNet and a neural vocoder.

The encoder consists of an embedding layer that embeds the discrete phonemes to a continuous space, three 1-D convolution layers followed by batch normalization and ReLU activations that add neighboring context information and a BLSTM layer that adds sequential information. To prevent overfitting, dropout is used in all convolution and LSTM layers with probability 0.5 and 0.1 respectively.

The decoder is an auto-regressive module, taking the output from the previous decoding step as the input of the current decoding step. It is composed of a pre-net and two Zoneout-LSTM \cite{Krueger2016ZoneoutRR} layers. To deal with exposure bias \cite{guo2019new}, we use Pre-net with two feedforward layers followed by dropout with a probability 0.5 to reduce the dependency on historical information. In each decoding step, we pass the output of the second LSTM layer to the attention module to get the corresponding vector from the encoder based on the weight calculated by the attention module. Finally, we use the attention output and the second LSTM output to predict the acoustic feature and the stop token. PostNet is a post-filter with five 1-D convolution layers to further improve the output quality. Conversational TTS often needs to synthesize long utterances which are easy to cause stability problems using location sensitive attention, so we adopt stepwise monotonic attention \cite{he2019robust}, which shows great  stability on long utterance generation. 

To reconstruct speech with high quality from the acoustic features, a neural vocoder \cite{kalchbrenner2018efficient,van2016wavenet} is usually used. In this work, we adopt Parallel Wavenet \cite{oord2017parallel} as our neural vocoder.

\begin{figure}[htp]
    \centering
    \includegraphics[width=0.45\textwidth]{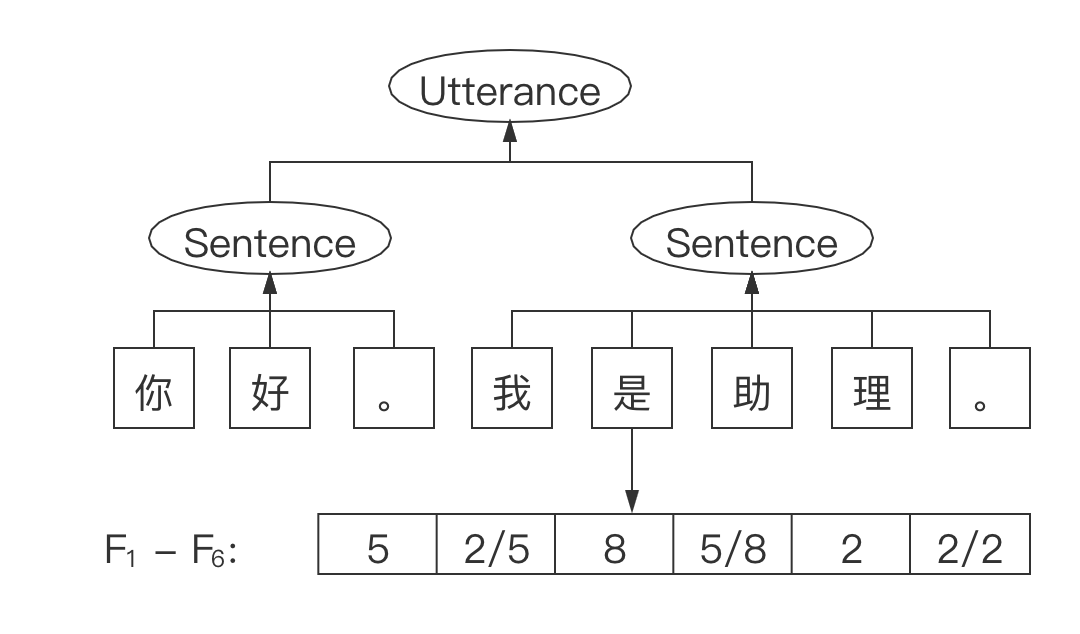}
    \caption{An example of extracting statistical features}
    \label{fig:sf}
\end{figure}

\begin{figure*}[htp]
    \centering
    \includegraphics[width=1.0\textwidth]{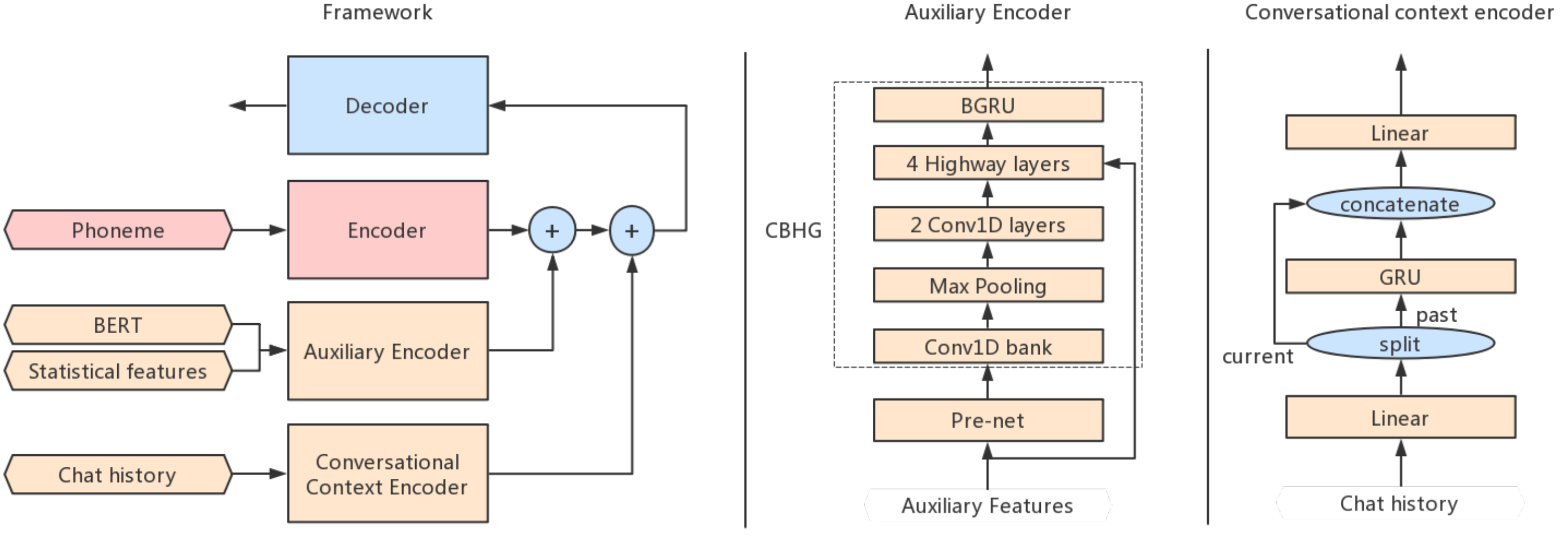}
    \caption{The framework of the conversational end-to-end TTS model. "+" donates the addition operation.}
    \label{fig:model}
\end{figure*}

\subsection{Auxiliary Encoder}

In addition to phonetic information, the text also contains rich information including semantics and syntax aspects, which affects prosody as well as naturalness. Auxiliary encoder is thus proposed to extract rich text features from these aspects to help produce more natural prosody. Word embedding is a kind of text representation, which has been previously used in SPSS \cite{Wang2015WordEF} and end-to-end TTS \cite{Ming2019FeatureRW} to help improve prosody by providing latent semantic and syntactic information learned from a rich repositery of texts. As the state-of-the-art word/sentence representation, BERT embedding \cite{Devlin2018BERTPO} has also shown its great performance to TTS \cite{xiao2020improving}.

Considering the fact that an utterance may be composed of more than one sentence, we also design a simplified group of statistical features to represent the syntactic structure. As shown in Figure \ref{fig:sf}, the features include the following items:
\begin{itemize}
    \item $F_1$: the number of characters in the current sentence.
    \item $F_2$: the relative-position of the current character in the current sentence.
    \item $F_3$: the number of characters in the current utterance.
    \item $F_4$: the relative-position of the current character in the current utterance.
    \item $F_5$: the number of sentences in the current utterance.
    \item $F_6$: the relative-position of the current sentence in the current utterance.
\end{itemize}
For Chinese, each ``character" is pronounced as a ``syllable" and all the integer values of statistical features are divided by a global max value to normalize to 0-1.

Auxiliary encoder uses BERT embeddings and the above statistical features to represent each character in a sentence. The encoder consists of a pre-net and a CBHG module, which has shown good modeling capability in Tacotron1 \cite{wang2017tacotron} (see the middle part of Figure \ref{fig:model}). The pre-net, composed of two linear layers with dropout rate 0.5, encodes the input firstly. The CBHG, composed of a group of CNN layers, highway network and GRU layer, can extract effective high-level sequential information to the TTS model. The features will be encoded by it, and up-sampled from character-level features to phoneme-level by replication, finally combined with the encoder outputs using the addition operation.

\subsection{Conversation Context Encoder}
\label{sec:context encoder}

In conventional conversational TTS, speech or dialogue act is often used as the conversation-related feature, which is either labeled manually, or predicted \cite{sridhar2011enriching}. However, this approach apparently has some problems. Firstly, the prediction error may lead to unnatural prosody when the wrong label is tagged as the input feature of the TTS model. Secondly, although the simple classification of sentences can represent various sentence patterns, it also inevitably results in the loss of important semantic and syntactic information.

\begin{figure}[htp]
    \centering
    \includegraphics[width=0.45\textwidth]{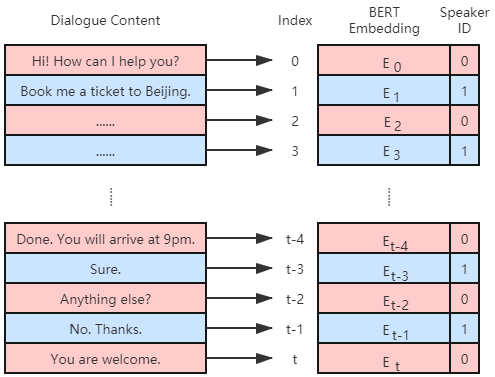}
    \caption{An example of the representation of the chat history. The left up-down block sequence donates the chat history. The utterance at index $t$ corresponds to a vector $H_t$, which is composed of a BERT embedding $E_t$ and a speaker ID.}
    \label{fig:chat_history}
\end{figure}

Therefore, we propose a conversational context encoder to directly extract prosody-related features from sentence embeddings with richer information rather than using sentence tags. Figure \ref{fig:chat_history} shows our approach to represent the chat history. As BERT can be used to extract both word embedding and sentence embedding, thus we still use it to extract sentence representation. And each embedding is attached by a one-hot vector as speaker ID to distinguish the customer and the agent.

The right part of Figure \ref{fig:model} shows the structure of the conversational context encoder. The conversational context encoder firstly processes the sequence of sentence embeddings $E_{t-c:t}$ with the length $c$ from $E_{t-c}$ to the current sentence $E_t$ through a linear layer\footnote{c includes both customer and agent text.}. We assume that text-related prosody is influenced by the current text and the cumulative state vector only determined by the past. So the chat history is divided into two parts, the current sentence $E_{t}$ and the past $E_{t-c:t-1}$. The GRU layer is used to encode the sequence $E_{t-c:t-1}$ to a state vector $S_{t}$ as the representation of the past. Then we concatenate $S_{t}$ and $E_t$, and feed them to the linear output layer to get the embedding vector of the sentence in a conversation. In this work, the length $c$ is set as the capacity of the chat history to abandon the old history which have less effect on the prosody of the current utterance.

\begin{table}[htp]
    \caption{The signal analysis parameters for audio preprocessing and Mel-spectrogram extraction. We firstly pre-process the audio using pre-emphasis, then extract the log-magnitude spectrogram via STFT analysis, finally get Mel-spectrogram and normalize it.}
    \centering
    \vspace{0.1cm}
    \begin{tabular}{c|c} \hline
    Pre-emphasis                   & coefficient: 0.97                      \\ \hline
    \multirow{2}{*}{STFT}      & Frame shift: 12.5ms                    \\ \cline{2-2}
                                   & Frame length: 50ms                     \\ \cline{2-2}
                                   & FFT size: 2048                         \\ \hline
    \multirow{2}{*}{Mel Spectrogram} & Dimension: 80    \\ \cline{2-2}
                                   & Min-max normalization: {[}-4, 4{]}              \\ \hline
    \end{tabular}
    \label{tab:stft}
\end{table}

\section{Experiments}

\subsection{Training Strategy \& Setup}

Our conversational corpus is recorded by two native Chinese female speakers who play the customer and the agent respectively. The original recording is the whole dialog track with the customer and the agent data together, and we processed the data into utterances. In this work, we only use the agent speech data containing about 2,000 utterances (3 hours) to train the TTS model for the voice agent. The conversational script is from real customer service conversations covering various scenarios, such as travel, training, sales, bank, etc..., to ensure the richness of context and scene. But the small training data size is still a challenge to train a highly stable end-to-end TTS model. Hence the encoder and decoder are firstly pre-trained with a standard TTS corpus, containing 6 hours of Chinese reading-style speech recorded by a native Chinese female speaker, to compensate for this problem.

There are three models used in the subjective evalution:
\begin{itemize}
    \item $M_1$: baseline mentioned in Section \ref{sec:baseline},
    \item $M_2$: $M_1$ plus auxiliary encoder,
    \item $M_3$: $M_2$ plus conversational context encoder.
\end{itemize}

\begin{table}[htp]
    \caption{The dimension of each layer in conversational context encoder}
    \centering
    \vspace{0.1cm}
    \begin{tabular}{c|c} \hline
    Layer & Dimension \\ \hline
    Input layer & 768 + 1 \\ \hline
    Input linear & 64 \\ \hline
    GRU & 64 \\ \hline
    Output linear & 512 \\ \hline
    \end{tabular}
    \label{tab:cce}
\end{table}

For all TTS models, the phoneme sequence contains phonemes, punctuations, inter-word and inter-syllable symbols. The output is Mel Spectrogram extracted from the recording audio with sample rate 16,000. Its specific analysis parameters are shown in Table\ref{tab:stft}.

$M_1$ is the baseline model which has the same structure and model parameter as Tacotron2 \cite{shen2018natural}, except for the attention module, which adopts stepwise monotonic attention (soft version) \cite{he2019robust} to improve stability on long utterance generation. $M_2$ is to verify the effect of introducing the auxiliary encoder, the pre-net and CBHG in the auxiliary encoder are the same as the corresponding modules in Tacotron1 \cite{wang2017tacotron}. We extract the character-level and utterance-level BERT embedding for each utterance individually. The character-level 768-dim BERT embedding and 6-dim statistical feature are replicated to phoneme level, and further encoded to 512-dim features by the auxiliary encoder, and combined with the encoder outputs using the addition operation. $M_3$ introduces the conversational context encoder with the configuration shown in Table\ref{tab:cce} to encode the 768-dim utterance-level BERT embedding and 1-dim speaker ID of the chat history $H_{t-c:t}$ mentioned in Section \ref{sec:context encoder}. The hidden layers are designed with a small structure to fit our corpus size. They can also be larger if the training set is large enough. We empirically set $c$ to 10, which shows better prosody than other value in our experiments.

All of the three models are trained on a single GPU with a batch size of 32. We use Adam optimizer with $\beta_1=0.9$, $\beta_2= 0.999$ and the learning rate exponentially decayed from $10^{-3}$ to $10^{-5}$ after 50,000 iterations to update these models.

\subsection{Subjective Evaluations}
\label{sec:subjective}

The test set is composed of five complete conversations between customer and agent (totally 31 conversation turns). The customer voice is recorded by a native Chinese male speaker and the agent part is synthesized using our models. A group of 20 Chinese native speakers takes part in comparison mean opinion score (CMOS) listening test, and they need to compare the prosodic performance according to the conversational context (Is prosodic expression more suitable and more natural with the current context?), and then rate on a scale from -3 to 3 with 1 point discrete increments according to the criteria shown in Table \ref{tab:cmos_s}. Firstly, these conversation turns are played one by one, and listeners rate every audio from the agent. Then we play the complete conversations again uninterruptedly, and listeners rate them based on the overall performance of the agent. Note that the customer voice is not subject to human scoring. The audio samples can be found at \url{https://hhguo.github.io/DemoConvTTS}.

\begin{table}[htp]
    \caption{CMOS scoring criteria}
    \centering
    \vspace{0.1cm}
    \begin{tabular}{c|c} \hline
    Score & Explanation (A v.s. B) \\ \hline
    -3 & A is much better \\ \hline
    -2 & A is better \\ \hline
    -1 & A is slightly better \\ \hline
     0 & About the same \\ \hline
     1 & B is slightly better \\ \hline
     2 & B is better \\ \hline
     3 & B is much better \\ \hline
    \end{tabular}
    \label{tab:cmos_s}
\end{table}

Experimental results in Table \ref{tab:cmos} show that auxiliary encoder improves the performance over the baseline model by CMOS score 0.22 and preference 42.9\% on the utterance level. On the conversation level, it also achieves superior performance with CMOS score 0.62 and preference 59.0\%. Compared with $M_2$, conversation context encoder can further improve the prosody expression by CMOS score 0.18 and preference 42.1\% on the utterance level, and CMOS score 0.39 and preference rate 57.0\% on the conversation level. We find that adding the auxiliary encoder can enhance the prosody expressiveness in term of stress, pause, and intonation by introducing semantic and structure features, and the conversation context encoder can make the agent response more natural and more consistent with the context. It shows that the context features provide richer prosody-related features.

\begin{table}[htp]
\caption{The results of CMOS tests. ``U-level" and ``C-level" donate the utterance level and conversation level respectively. Preference is calculated according to CMOS scores: the score greater than 0 means bias towards B, equal to 0 means neutral and less than 0 means bias towards A.}
\centering
\vspace{0.1cm}
\begin{tabular}{cccccc} \hline
 & \multirow{2}{*}{CMOS} & \multicolumn{4}{c}{Preference (\%)} \\ \cline{3-6} 
 &      & $M_1$ & Neutral & $M_2$ & $p$-value \\ \hline
U-level & 0.22 & 24.4 & 32.7 & \textcolor{red}{42.9} & 0.0001 \\ \hline
C-level & 0.62 & 21.0 & 20.0 & \textcolor{red}{59.0} & 0.0001 \\ \hline
\vspace{0.1cm}
\end{tabular}

\begin{tabular}{cccccc} \hline
 & \multirow{2}{*}{CMOS} & \multicolumn{4}{c}{Preference (\%)} \\ \cline{3-6} 
 &      & $M_2$ & Neutral & $M_3$ & $p$-value \\ \hline
U-level & 0.18 & 28.1 & 29.8 & \textcolor{red}{42.1} & 0.0001\\ \hline
C-level & 0.39 & 28.0 & 15.0 & \textcolor{red}{57.0} & 0.001 \\ \hline
\end{tabular}
\label{tab:cmos}
\end{table}

\subsection{Spontaneous Behaviors}
\label{sec:spontaneous}

In our corpus, most of spontaneous behaviors are about hesitation as mentioned in Section \ref{sec:Spontaneous}. Note that, we do not need any extra label of them, and just make conversation end-to-end TTS model learn to express these behaviors by itself. By inserting the appropriate spontaneous behaviors to the synthesized content, we find that our models have the ability to properly express them, such as fillers, repeated word, which can make the conversational speaking style more realistic to listeners.

\section{Conclusion}

In this work, we firstly design a well-recorded spontaneous conversational TTS corpus for voice agent. This corpus adopts a new recording approach, recorded in three steps: scene and dialogue design, recording in the form of performance and transcribing. To produce high-quality conversational speech for voice agent, we then propose a conversational context-aware end-to-end TTS approach. Specifically, we add an auxiliary encoder to extract rich semantics and syntax information to help produce more natural prosody, and a conversational context encoder to improve awareness on the contextual information. Experimental results show that both of these two encoders are effective to help the voice agent to produce better prosody in the conversation, and some spontaneous behaviors can be well expressed by the proposed end-to-end models.

\bibliographystyle{IEEEtran}

\bibliography{refs}

\end{document}